\documentstyle[epsf,epsfig]{aipproc2}

\newcommand{\C}{{\cal C}}
\newcommand{\calN}{{\cal N}}
\newcommand{\calD}{{\cal D}}
\newcommand{\muK}{\mu{\rm K}}

\def\tN{{\widetilde N}}
\def\tA{{\widetilde A}}

\def\td{{\tilde d}}
\def\ts{{\tilde s}}

\def\teps{{\tilde\varepsilon}}
\def\cF{{\cal F}}
\def\cG{{\cal G}}
\def\bN{{\bar N}}
\def\bT{{\bar T}}

\begin{document}
\title{CMB Likelihood Functions for Beginners and Experts}

\author{Andrew H. Jaffe$^1$, J.R. Bond$^2$, P.G. Ferreira$^{3,4}$,
  \& L.E. Knox$^5$} 
\address{$^1$Center for Particle Astrophysics, University of California, 
  Berkeley, CA 94720,\\
$^2$CITA, Toronto, ON, Canada \\
$^3$Theory Group, CERN, CH-1211, Geneve 23, Switzerland\\ 
$^4$CENTRA, Instituto Superior Tecnico, Lisboa 1096 Codex, Portugal\\
$^5$Astronomy \& Astrophysics Center, U.~Chicago, Chicago, IL }

\lefthead{Jaffe et al.}
\righthead{CMB Likelihood Functions}

\maketitle

\begin{abstract}
  Although the broad outlines of the appropriate pipeline for
  cosmological likelihood analysis with CMB data has been known for
  several years, only recently have we had to contend with the full,
  large-scale, computationally challenging problem involving both
  highly-correlated noise and extremely large datasets ($N > 1000$). In
  this talk we concentrate on the beginning and end of this process.
  First, we discuss estimating the noise covariance from the data itself
  in a rigorous and unbiased way; this is essentially an iterated
  minimum-variance mapmaking approach. We also discuss the unbiased
  determination of cosmological parameters from estimates of the power
  spectrum or experimental bandpowers.
\end{abstract}

\section{Introduction}

In principle, CMB anisotropy data analysis is easy: for at least some
subset of theories (i.e., inflation) and some subset of CMB experiments,
exact likelihood functions can be written down. In these cases, at
least, the rest of the problem is implementation: the numerical
calculation of the probability distribution of the interesting
parameters given the data.

We model a measurement of the CMB as
\begin{equation}
  \label{eq:data}
  d_i = \sum_p A_{ip} T_p + n_i
\end{equation}
where $d_t$ is the data at time $t_i$, $T_p$ is the beam-smeared
temperature at pixel $p$ (pointed at by unit vector ${\bf\hat x}_p$),
$A_{ip}$ is the pointing operator such that $A_{ip}=1$ when pixel $p$ is
observed at time $i$, and $n_i$ is the noise at time $i$. The noise is
characterised by its distribution, for our purposes assumed to be a
Gaussian with covariance $\langle n_i n_{i'} \rangle=N_{ii'}$

Given the data, then, we first wish to solve for the sky temperature
$T_p$ and its posterior distribution, and then use this map to determine 
the statistical properties of the CMB sky, such as its power spectrum,
$C_\ell$ and its distribution, from which we will finally determine the 
underlying cosmological parameters.

Unfortunately, this implementation is indeed complicated in realistic
cases. It is usually assumed that the error distribution for an
experiment is somehow handed down to the theorists and data analysts by
the experimenters, themselves with a direct link to some higher being
who decides these things. Of course this is not actually the case:
instead, the error distribution itself must be estimated from the data
along with the signal. In some (not necessarily realistic) cases, this
is simple. If it were known {\em a priori} that the noise were Gaussian
and white, then $N_{ii'}=\sigma^2 \delta_{ii'}$ and we could use the
histogram of observations at a given pixel to estimate the noise
variance $\sigma$.

In more realistic cases, the noise in the timestream will be correlated
and contaminated by systematic problems and glitches such as cosmic
rays, making any simple procedure like this infeasible in practice. In
this paper we outline a self-consistent Bayesian (likelihood) method for 
determining the noise power spectrum along with the CMB map.

Given a map and its statistical distribution (i.e., the error matrix for 
Gaussian noise), determining the power spectrum (for the case of a
Gaussian CMB signal, as well) has been well-studied; problems remain in
the implementation of efficient algorithms, but the ideal solution is
known.

In this paper, we also address the question of what to do next: how do
we go from an estimate of the CMB fluctuation power spectrum, $C_\ell$,
to the underlying cosmological parameters? Because of the complicated
(non-Gaussian) nature of the likelihood function, we will show that
simple $\chi^2$ techniques will necessarily bias the determination of
the parameters, but will derive an ansatz for eliminating this bias.

The plan of this paper is as follows. First, we discuss the CMB
Data-Analysis Pipeline from the standpoint of likelihood functions. Then 
we address the specific problem of simultaneously determining the noise
and signal from the timestream. Finally, we address the problem of
estimating parameters from the power spectrum.

\section{Likelihoods for CMB Experiments}

As has become customary, we start our analysis with Bayes' theorem
\begin{equation}
  \label{eq:bayestheorem}
  P(\theta|DI) \propto P(\theta|I) P(D|\theta I)
\end{equation}
where $\theta$ are the parameters we are trying to determine, $D$ is the 
data, and $I$ is the ``background information'' describing the
problem. The quantity $P(D|\theta I)$ is thus the likelihood, the
probability of the data given a specific set of parameters,
$P(\theta|I)$ is the prior probability for the parameters, and the
left-hand side of the equation is the posterior probability for the
parameters given the data.

As above and throughout, we will take the data, $d_i$, as given by a
time series of CMB measurements,
\begin{equation}
  \label{eq:timeseries}
  d_i = s_i + n_i = \sum_p A_{ip} T_p + n_i,
\end{equation}
where $i$ labels the time, $t=i\delta t$, $s_i$ and $n_i$ are the
experimental noise and sky signal contributions at that time. The signal
is in turn given by the operation of a ``pointing matrix,'' $A_{ip}$ on
the sky signal at pixel $p$, $T_p$ (i.e., the ``map''); we take the
latter to be already pixelized and smeared by the experimental beam, so
$A$ is a very sparse matrix with a single ``1'' entry for each time
corresponding to the observed pixel. In the future we will often rely on 
the summation convention and write $\sum_p A_{ip} T_p = A_{ip}T_p$, or
occasionally use matrix notation, as in $AT$, etc.

We will assume that the observed noise $n_i$ is a realization of a
stationary Gaussian process with power spectrum $\tN(\omega)$. This means
that the correlation matrix of the noise is given by
\begin{equation}
  \label{eq:noisecorrelation}
  \langle n_i n_{i'} \rangle = N_{ii'} = \int {d\omega\over2\pi}\;
  \tN(\omega) e^{-i\omega \delta t(i-i')}.
\end{equation}
The stationarity of the process requires (or is defined by) $N_{ii'} =
N(t_i-t_i')$.

Most generally, we will take the parameters to be
\begin{itemize}
\item The observed CMB signal on the sky, $T_p$;
\item The power spectrum of the noise, $\tN(\omega)$
\item (Possibly) any cosmological parameters which describe the
  distribution of the $T_p$ (i.e., the CMB power spectrum, $C_\ell$,
  although we could also use the cosmological paramters such as $H_0$
  and $\Omega$ directly).
\end{itemize}
Sometimes, we will assign a prior distribution for the CMB signal such
that the cosmological parameters will be irrelevent; other times we will 
marginalize over the CMB signal itself and determine those parameters.

With these parameters and the data, $d_i$, Bayes' theorem becomes
\begin{equation}
  \label{eq:bayescmb}
  P[T_p,\tN,C_\ell | d_i,I] \propto
  P[\tN|I] \; P(T_p,C_\ell | I) \;
 P[d_i | \tN(\omega), T_p, I].
\end{equation}
Here, we have used two pieces of information to simplify slightly.
First, the noise power spectrum, $N(\omega)$ does not depend at all on
the signal, so we can separate out its prior distribution. Second, given
the noise power spectrum and the sky signal, the likelihood does not
depend upon the cosmological parameters. For the Gaussian noise we
assume, the likelihood is simply
\begin{eqnarray}
  \label{eq:noiselike}
  -2\ln{\cal L}&\equiv&-2\ln P[d_i | \tN, T_p]  \nonumber\\
  &=& \ln \left| N_{ii'} \right| +
  \sum_{ii'} (d_i-s_i)N^{-1}_{ii'}(d_{i'}-s_{i'}) =
 \sum_k \left[ \ln \tN_k +
    |\td_k-\ts_k|^2/\tN_k \right]
\end{eqnarray}
(ignoring an additive constant); recall that $s_i=\sum_p A_{ip}T_p$.
The second equality uses tildes to denote the discrete Fourier transform
at angular frequency $\omega_k$ (see appendix and/or below\ldots).

We will now apply these general formulae to various cases.

\subsection{Known noise power spectrum}
\label{sec:knownnoise}
We will start with the simplest case, where we have complete prior
knowledge of the noise power spectrum. This is the case that has been
previously discussed in the literature, but we emphasize that it is very
unrealistic!

We assign a delta-function prior distribution to $N$, transforming it in
effect from a parameter to part of the prior knowledge. First, we assume 
no cosmological information about the distribution of temperatures on
the sky: $P(T_p, C_\ell | I ) = P(T_p | I) P(C_\ell | I)$; with this
separation the posterior for $C_\ell$ is simply the prior---the
experiment gives us no new information. We will also assign a uniform
prior to $T_p$, lacking further information. Now, the posterior
distribution for the sky temperature is simply proportional to the
likelihood, which can be rewritten by completing the square as
\begin{equation}
  \label{eq:maplike}
  P(T_p | \tN, d, I) \propto P(d|T_p,\tN,I)
=
 {1\over\left| 2\pi C_{Npp'} \right|^{1/2}}
\exp\left[ -{1\over2} \sum_{pp'} \left(T_p - \bT_p\right) C^{-1}_{N,pp'}
    \left(T_{p'} - \bT_{p'}\right)\right]
\end{equation}
with the mean (also, the likelihood maximum) given by
\begin{equation}
  \label{eq:lsmap}
  \bT = \left( A^T N^{-1} A\right)^{-1} A^T N^{-1} d
\end{equation}
(in matrix notation), and the noise correlation matrix by
\begin{equation}
  \label{eq:noisecorr}
  C_N = \left( A^T N^{-1} A\right)^{-1}.
\end{equation}
Occasionally, the inverse of this correlation matrix is referred to as
the {\em weight matrix}.
As is usual for linear Gaussian models, the mean is just the
multidimensional least-squares solution to $d = A T$ with noise
correlation $N$.  This is just the mapmaking procedure advocated in
Refs.~\cite{wrightprocedure,tegmaps}, cast into the form of a
Bayesian parameter-estimation problem.

For the case of known noise, however, this map is more than a just a
visual representation of the data. Even if we wish to determine the
cosmological parameters, it is an essential quantity. We can write
the prior for both the map and the spectrum as
\begin{equation}
  \label{eq:sigprior}
  P(C_\ell, T_p | I) = P(T_p | C_\ell, I) P(C_\ell | I)
\end{equation}
using the laws of probability, and so we can see that our rewriting of
the likelihood in the form of Eq.~\ref{eq:maplike} remains useful. That
is, the full distribution is only a function of the data through the
maximum-likelihood map, $\bT$---in statistical parlance, $\bT$
is a {\em sufficient statistic}.  Thus for known noise, we can {\em
  always} start by making a map (and calculating its noise matrix,
$C_N$).

\subsubsection{Cosmological CMB priors}
\label{sec:cosmopriors}
Here, we briefly examine the specific form of the signal prior, $P(T_p |
C_\ell, I)$, motivated by simple Gaussian models. That is, we take the
sky temperature, $T_p$, to be an actual realization of a Gaussian CMB
sky, with covariance specified by the power spectrum, $C_\ell$,
\begin{equation}
  \label{eq:signalcovar}
  \langle T_p T_{p'} \rangle = C_{T,pp'} =
  \sum_\ell\frac{2\ell+1}{4\pi}C_\ell B_\ell^2 P_\ell
  \left({\hat x}_p\cdot{\hat x}_{p'}\right)
\end{equation}
(note that we include beam-smearing by a symmetric beam with spherical
harmonic transform $B_\ell$ in this definition); ${\hat x}_p\cdot{\hat
  x}_{p'}$ gives the cosine of the angle between the pixels. With this
covariance, the prior becomes
\begin{equation}
  \label{eq:signalprior}
  P(T_p |C_\ell, I) = {1\over\left| 2\pi C_{Tpp'} \right|^{1/2}}
  \exp\left[ -{1\over2} \sum_{pp'} T_p C^{-1}_{T,pp'} T_{p'}\right].
\end{equation}
We thus have a posterior distribution for $T_p$ and $C_\ell$ which is
the product of two Gaussians. In the usual cosmological likelihood
problem, we don't care about the actual sky temperature {\em per se},
but are concerned with the $C_\ell$ (or the parameters upon which the
power spectrum depends). Thus, we can marginalize over the $T_p$,
\begin{eqnarray}
  \label{eq:mapmarginalize}
  P(C_\ell | d, I) &=& \int dT_p\; P(C_\ell,T_p|d,I) = 
P(C_\ell|I) \int dT_p\; P(T_p|C_\ell,I) P(d|T_p,I) \nonumber =
P(C_\ell|I) P(\bT(d)|C_\ell I)\\
&=& P(C_\ell|I)
{1\over \left| 2\pi C_T \right|^{1/2}}  {1\over \left| 2\pi C_N \right|^{1/2}}
\exp\bigg[-{1\over2} \sum_{pp'}
\left( T_p C^{-1}_{T,pp'} T_{p'} +
      \left(T_p - \bT_p\right) C^{-1}_{N,pp'} 
      \left(T_{p'} - \bT_{p'}\right) \right)\bigg]
\end{eqnarray}
where we have incuded a prior for the power spectrum itself, so we can
write the Gaussian factor as the likelihood for the map given $C_\ell$,
$ P(C_\ell | d, I)\propto P(C_\ell|I)P(\bT|C_\ell)$.
The equation defines the effective likelihood for the map ($\bT$, now
considered as the data rather than the timestream itself, $d$), which is
easily computed again by completing the square, giving
\begin{equation}
  \label{eq:maplikelihood}
  P(\bT_p | C_\ell I) =
  {1\over\left| 2\pi\left(C_{Tpp'} + C_{Npp'}\right)\right|^{1/2}}
\exp\left[ -{1\over2} \sum_{pp'} \bT_p \left(C_T+C_N\right)^{-1}_{pp'}
    \bT_{p'}\right].
\end{equation}
This is just the usual CMB likelihood formula: the ``observed map,''
$\bT_p$, is just the sum of two quantities (noise and signal)
distributed as independent gaussians. Note again that the data only
enters through the maximum likelihood map, $\bT$, although that
calculation is only implicit in this formula. Further, the power
spectrum $C_\ell$ only enters through the signal correlation matrix,
$C_T$, and in a very nonlinear way.

We can also play a slightly different game with the likelihood. If we
retain the gaussian prior for the CMB temperature but {\em fix} the CMB
power spectrum, we can estimate the map with this additional prior
knowledge. We will again be able to complete the square in the
exponential and see that $T_p$ is distributed as a gaussian:
\begin{equation}
  \label{eq:wienerlike}
  P(T_p | C_\ell, d, I)={1\over \left| 2\pi C_W \right|^{1/2}} 
  \exp\left[-{1\over2} \chi^2(T_p | C_\ell, d, I)\right]
\end{equation}
with
\begin{equation}
  \chi^2(T_p | C_\ell, d, I)\equiv\sum_{pp'} \
      \left(\bT_p - (W\bT)_p\right) C^{-1}_{W,pp'} 
      \left(\bT_{p'} - (W\bT)_{p'}\right).
\end{equation}
Now, the mean is given by
\begin{equation}
  \label{eq:wiener}
  (W\bT) = C_T(C_T + C_N)^{-1}\bT
\end{equation}
which is just the {\em Wiener Filter}! It has correlation matrix given
by
\begin{equation}
  \label{eq:wienercorr}
  C_W = C_T(C_T + C_N)^{-1}C_T.
\end{equation}
Note that the maximum-likelihood map, $\bT$ still appears in these
formulae, but it is no longer the maximum of the {\em posterior}
distribution, now given by the Wiener filter, $W\bar T$.

This subsection has shown how many of the usual CMB data calculations
can be seen as different uses of the Bayesian formalism:
\begin{itemize}
\item the least-squares map (seeing it as a ``sufficient statistic''),
\item the CMB cosmological-parameter likelihood function, and
\item the Wiener filter map of the CMB signal.
\end{itemize}
The differences depend on what quantity is estimated (the map or the
power spectrum) and what prior information is included.

\section{Simultaneous Estimation of the Noise and Sky Map}

We now turn to the (realistic!) case where the noise covariance of the
experiment are not known prior to receiving the data. That is, we must
use the data itself to estimate the noise properties.

The previous subsection briefly outlined the Bayesian approach to CMB
statistics, assuming a known noise power spectrum. Now, we will relax
this assumption and approach the more realistic case when we must
estimate both the experimental noise and the anisotropy of the CMB. We
will take as our model a noise power spectrum of amplitude $\bN_\alpha$
in bands numbered $\alpha$, with a shape in each band given by a fixed
function $P_k$ with a width of $n_\alpha$ (the number of discrete modes
in the band); we will usually take $P_k={\rm const}$ so $\tN$ is
piecewise constant. That is,
\begin{equation}
  \label{eq:noisemodel}
  \tN(\omega_k) = \bN_\alpha P_k.
\end{equation}

We again assign a constant prior to the sky map, $T_p$. As the prior for
the noise we will take $P(\bN_\alpha)\propto1/\bN_\alpha^\nu$
With $\nu=1$ and a single band, this is the usual Jeffereys prior
advocated for ``scale parameters''  and the units on ${\bar
  N}_\alpha$ are irrelevent.

With this model and priors, the joint likelihood for the noise and the
map becomes
\begin{eqnarray}
  \label{eq:jointprob}
  P(T_p, \bN_\alpha | d,I) &\propto& \prod_\alpha
  {1\over\bN_\alpha^{\nu+n_\alpha/2}} 
  \exp\left[ -{1\over2\bN_\alpha}
    \sum_{k\in\alpha}\frac{1}{P_k}\left|\td_k - \tA_{kp}T_p\right|^2\right]
\nonumber\\
  &=&  \prod_\alpha
  {1\over\bN_\alpha^{\nu+n_\alpha/2}} 
  \exp\left[ -{1\over2\bN_\alpha}
    \sum_{k\in\alpha}\frac{1}{P_k}\left|\teps_k\right|^2\right]
\end{eqnarray} 
where ${k\in\alpha}$ refers to a sum over modes in band $\alpha$. We
also define the estimate of the noise as $\varepsilon=d-AT$ for future use.

We can simultaneously solve for the maximum-probability noise and
signal. Carrying out the derivatives gives
\begin{eqnarray}
  \label{eq:Tpderiv}
  {\partial \ln{\cal L}\over\partial T_p} &=& \sum_\alpha\left[
    {1\over\bN_\alpha}\sum_{k\in\alpha} \frac{1}{P_k}
    \left(d_k-A_{kp'}T_{p'}\right)A_{kp}\right]\nonumber\\
&=& (d-AT)^T N^{-1} A = \varepsilon^T N^{-1} A
\end{eqnarray}
(switching between indices in Fourier space and matrix notation) and 
\begin{equation}
  \label{eq:Nderiv}
  {\partial \ln{\cal L}\over\partial \bN_\alpha} = -{1\over2}
  {1\over\bN_\alpha}\left[(n_\alpha+2\nu)-{1\over\bN_\alpha}
      \sum_{k\in\alpha} 
      \frac{1}{P_k}\left|\td_k-\tA_{kp'}T_{p'}\right|^2\right]
\end{equation}
Setting these to zero and solving, we then find me must simultaneously
satisfy
\begin{equation}
  \label{eq:maxLmap}
  T = (A^T N^{-1} A)^{-1} A^T N^{-1} d
\end{equation}
(using matrix notation for simplicity) and
\begin{equation}
  \label{eq:maxLnoise}
  \bN_\alpha = {1\over n_\alpha+2\nu}\sum_{k\in\alpha}
  \frac{1}{P_k}\left|\td_k-\tA_{kp'}T_{p'}\right|^2 = 
 {1\over n_\alpha+2\nu}\sum_{k\in\alpha}
  \frac{1}{P_k}\left|\teps_k\right|^2.
\end{equation}
Equation~\ref{eq:maxLmap} is just the usual maximum-likelihood map
solution; Equation~\ref{eq:maxLnoise} is the average ``periodogram'' of the 
noise over the band $\alpha$, with a slight modification for the prior
probability, parameterized by $\nu$; for wide bands this is irrelevant.
As we will see below, iteration is actually a very efficient way to solve these
simulataneously. For future reference, we also write down the derivative 
with respect to the map at the joint maximum (i.e., substituting
Eq.~\ref{eq:maxLnoise} into Eq.~\ref{eq:Tpderiv}),
\begin{equation}
  \label{eq:maxLTpderiv}
    {\partial \ln{\cal L}\over\partial T_p} =\sum_\alpha\left[
      \left(n_\alpha+2\nu\right){
   \sum_{k\in\alpha} \teps_k\tA_{kp}/P_k
\over
   \sum_{k'\in\alpha} \left|\teps_{k'}\right|^2/P_{k'}}
      \right]
\end{equation}

If we fix the noise at the joint maximum, then the problem reduces to
that of the previous section, a Gaussian likelihood in $T_p$ with which
the usual tools can be applied. This is not a rigorous approach to the
problem, however due to correlation between noise and signal
estimation. To get a handle on this, we calculate the curvature of the
distribution around this joint maximum, which we define as
\begin{equation}
  \cF = \left(
    \begin{array}{cc}
      \cG_{pp'} & \cG_{kp} \\
      \cG_{pk} & \cG_{kk'} \\
    \end{array}\right)
\end{equation}
where a subscript $\alpha$ or $p$ refers to a derivative with respect to
$\bN_\alpha$ or $T_p$, respectively.  $\cF$ refers to the full matrix;
$\cal G$ to the sub-blocks. Explicitly, the parameter derivatives are
\begin{equation}
  \label{eq:maxcurvTT}
  {\partial^2 \ln{\cal L}\over\partial T_p\partial T_{p'}}=\cG_{pp'} = 
-\sum_\alpha \frac{1}{\bN_\alpha}\sum_{k\in\alpha}\frac{\tA_{kp}\tA_{kp'}}{P_k}
=-A^T N^{-1} A
\end{equation}
and
\begin{equation}
  \label{eq:maxcurvNN}
  {\partial^2 \ln{\cal L}\over\partial \bN_\alpha\partial
    \bN_{\alpha'}}=
  \cG_{\alpha\alpha'}=-\frac{n_\alpha+2\nu}{2\bN_\alpha^2}
  \delta_{\alpha\alpha'}
\end{equation}
with the cross-curvature
\begin{equation}
  \label{eq:maxcurvTN}
  {\partial^2 \ln{\cal L}\over\partial T_p\partial \bN_{\alpha}}=
\cG_{p\alpha}=-\frac{1}{\bN_\alpha^2}
\sum_{k\in\alpha}\frac{\teps_k \tA_{kp}}{P_k}
\end{equation}

The distribution about the joint maximum is {\em not} a Gaussian in the
$\bN_\alpha$ directions, so there is more information available than
this. Nonetheless, if we treat the distribution as if it were Gaussian,
we can calculate the covariance matrix, given by the inverse of this
curvature matrix:
\begin{equation}
  \cF^{-1} = \left[\begin{array}{cc}
      \left(\cG_{pp'}-\cG_{p\alpha} \cG^{-1}_{\alpha\alpha'} \cG_{\alpha'p}\right)^{-1} &
      \left(\cG_{\alpha p} -\cG_{\alpha\alpha'}\cG^{-1}_{\alpha'p'}\cG_{p'p}\right)^{-1} \\
      \left(\cG_{p\alpha} -\cG_{pp'}\cG^{-1}_{p'\alpha'}\cG_{\alpha'\alpha}\right)^{-1} &
      \left(\cG_{\alpha\alpha'}-\cG_{\alpha p} \cG^{-1}_{pp'} \cG_{p'\alpha}\right)^{-1} 
    \end{array}\right]
\end{equation}
In particular, the effective variance of the map is increased from $C_N$
to 
\begin{eqnarray}\label{eq:effectivevar}
C^{\rm eff}_{N pp'} &=& (C_{N \ pp'}^{-1}-\cG_{p\alpha}\cG^{-1}_{\alpha\alpha'}\cG_{\alpha'p'})^{-1}\nonumber\\
&=&\left [\sum_\alpha \frac{1}{\bN_\alpha}\sum_{k\in\alpha}\frac{\tA_{kp}\tA_{kp'}}{P_k}- 2\sum_\alpha \frac{1}{(n_\alpha + 2\nu)\bN^2_\alpha}
  \sum_{k\in\alpha}\frac{\teps_k\tA_{kp}}{P_k}
  \sum_{k'\in\alpha}\frac{\teps_{k'}\tA_{k'p'}}{P_{k'}}\right ]^{-1}
\end{eqnarray}
which we have evaluated at the simultaneous peak of the distribution to
eliminate $\bN_\alpha$.

In Ref.~\cite{FJ99}, we examine this correction to $C_N^{-1}$ (the
weight matrix) more closely;
for sufficiently wide bands, relative to the number of pixels in the
map, the correction is negligeable.

\subsection{Noise Marginalization}

Instead of this joint solution, formally, at least, we know the
appropriate procedure: marginalize over the quantity we don't care about
(the noise power spectrum) to obtain the distribution for the quantity
we wish to know (the map, $T_p$). We can actually carry out this
integral in this case:
\begin{equation}
  \label{eq:noisemarg}
  P(T_p | d_i,I) =  \int d\tN_k\; P(T_p, N_k| d_i, I)
\propto \prod_\alpha\left[\sum_{k\in\alpha}
    \frac{1}{P_k}\left|\td_k - \tA_{kp}T_p\right|^2\right]^{-(n_\alpha/2+\nu-1)}
\end{equation}
(this is just Student's t distribution in a slightly different form than 
is usually seen).
If we stay in Fourier space, the maximum of this distribution can be
calculated to be the solution of
\begin{equation}
  \label{eq:margmax}
    {\partial \ln P \over\partial T_p} =\sum_\alpha\left[
       \left(n_\alpha+2\nu-2\right){
     \sum_{k\in\alpha} \frac{1}{P_k}\left(d_k-A_{kp'}T_{p'}\right)\tA_{kp}
\over
     \sum_{k'\in\alpha}\frac{1} {P_{k'}}\left|d_{k'}-\tA_{k'p'}T_{p'}\right|^2}
      \right]
\end{equation}
Note that this is {\em exactly the same form} as
Eq.~\ref{eq:maxLTpderiv}, the equation for the maximum probability
map in the joint estimation case, with the prefactor $(n_\alpha+2\nu)$
replaced by $(n_\alpha+2\nu-2)$. This
is equivalent to changing the exponent of the prior probability from
$\nu$ to $\nu-1$: the marginalized maximum for $\nu=1$ (Jefferys prior)
is the same as the joint maximum for $\nu=0$ (constant prior). We have
also seen that for $n_\alpha\gg1$, the value of $\nu$ is irrelevent, so
that these should be nearly equal. (Moreover, the numerical tools to
calculate the joint solution, outlined below, can be used in this case,
too.) Note also that if $n_\alpha+2\nu-2\le 0$, the equation isn't solved
for any (finite) map. In this case, either our prior information is so
unrestrictive or the bands are so narrow that it is impossible to
distinguish between noise and signal, and the probability distribution
has no maximum when the noise is marginalized.

In principle, any further analysis of the map would have to rely on the
full distribution of Eq.~\ref{eq:noisemarg}. In practice, the
t-distribution is quite close to a Gaussian [again, for
$n_\alpha\gg1$---is this right?] (although the tails are suppressed by a
power-law rather than an exponential). It will thus be an excellent
approximation to take the distribution to be a Gaussian
\begin{equation}
  \label{eq:noisemargapprox}
  P(T_p | d_i,I) \approx
  {1\over\left| 2\pi C_{N,pp'}^{\rm eff} \right|^{1/2}}
  \exp\left[ -{1\over2} \sum_{pp'} 
\left(T_p - \bT_p\right) \left(C^{\rm eff}_{N,pp'}\right)^{-1}
    \left(T_{p'} - \bT_{p'}\right)\right]
\end{equation}
with noise covariance given by
\begin{equation}
  \label{eq:CNeff}
  \left(C_{Npp'}^{\rm eff}\right)^{-1}=
  -{\partial^2 \ln P(T_p | d_i,I)\over\partial
    T_p\partial T_{p'}}
=\sum_\alpha \left[\frac{(n_\alpha+2\nu-2)}{\sum_{k\in\alpha}|\teps_k|^2}\sum_{k\in\alpha}\frac{\tA_{kp}\tA_{kp'}}{P_k}- 2 \frac{(n_\alpha+2\nu-2)}
{(\sum_{k\in\alpha}|\teps_k|^2)^2 }
  \sum_{k\in\alpha}\frac{\teps_k\tA_{kp}}{P_k}
  \sum_{k'\in\alpha}\frac{\teps_{k'}\tA_{k'p'}}{P_{k'}}\right ]
\end{equation}
where $\bT$ is the solution to Eq.~\ref{eq:margmax} and the derivatives
are taken with respect to the {\em full} distribution of
Eq.~\ref{eq:margmax}).  Note that the inverse of $C_N^{\rm eff}$ (the
weight matrix) is given again by the sum of two terms. The first is just
the weight matrix that would be assigned given the maximum probability
solution for the noise, Eq.~\ref{eq:noisecorr} (with the numerically
irrelevent change of the prior $\nu\to\nu-1$ as noted above). Again we
find a correction term due to the fact that we are only able to {\em
  estimate} this noise power spectrum, where the correction has exactly
the same form as in the non-marginalized case
(Eq.~\ref{eq:effectivevar}) above with a change of the prefactor.

In Ref.~\cite{FJ99}, we show how the simultaneous solution for noise and map
can be performed iteratively. As we have shown, this is also the
solution for the case when the noise is marginalized over.

\subsection{Discussion}
Happily, the bottom line of this analysis is that the mapmaking phase of 
CMB data analysis can proceed as has been discussed by other authors in
the past. We do recommend two slight additions to the pipeline:
\begin{itemize}
\item Iterative map \& noise determination
  \begin{enumerate}
  \item Start by assuming the data stream is all noise.
  \item Calculate the noise power spectrum via FFT.
  \item Inverse FFT, and use this noise matrix for mapmaking
  \item Subtract the estimated signal contribution from the map, and
    repeat.
  \end{enumerate}
\item Calculate the correction to the noise matrix (Eq.~\ref{eq:CNeff})
  (and either use the correction or make sure it is negligeable).
\end{itemize}

We note that it is only necessary to solve for the map itself during the 
iterative process, but not to invert the inverse noise matrix,
$C_N^-1$. The former can, in principle, be accomplished in far fewer
operations than a matrix inverse, so the iterative procedure need not
add a great amount of computing time to the process.

However, it is conventional wisdom that an ideal observational strategy has
signal to noise of unity.  In this regime, the signal to noise {\em in
  the timestream} is very low, and a good aproximation to the estimator
of the noise power spectrum is the banded periodogram of the data stream
itself. Thus one is able to avoid the time-consuming process of
performing multiple iterations.

\section{Radical Compression:\\ 
From the Power Spectrum to Cosmological Parameters}

Above, we discussed how to make a map of the CMB and estimate its noise
properties. Elsewhere, the estimation of the CMB anisotropy power
spectrum ($C_\ell$) from a map and its noise estimate have been
described and implemented at least for data sets of moderate size. But
the ultimate goal, of course, is to go from the power spectrum to the
cosmological parameters. Various groups \cite{paramest} have calculated
how well future experiments are expected to perform; here, we describe
what is necessary to begin to achieve such results.

The basic problem is that a plot power spectrum estimates and associated
errors do not suffice to describe the likelihood function of the
experiment. Usually, we plot $\C\equiv\ell(\ell+1)C_\ell/(2\pi) \pm
\sigma_\ell$ and are tempted to do a $\chi^2$ procedure using those
errors and possibly covarinaces between $\ell$ to fit theoretical curves
to the estimates. As we have seen above, however, the shape of the
likelihood as a function of the $C_\ell$ is not Gaussian, implicitly
assumed in the use of a $\chi^2$ procedure.

One solution is simply to use a full likelihood-minimization procedure
on the cosmological parameters given the entire dataset. That is,
evaluate Eq.~\ref{eq:maplikelihood} directly for points throughout the
cosmological parameter space (over which $C_\ell$ and thereby the
correlation matrix $C_T$ will vary).  Since all calculations of the likelihood
function to date require $O(N_{\rm pix}^3)$ operations in general, this
is prohibitive for experiments of even moderate size (especially if the
likelihood must be searched over something like the 11-dimensional space
appropriate for inflation-inspired models).

Instead, we must find efficient ways to approximate the likelihood
function based on minimal information. In the rest of this paper, we
discuss approximations motivated by our
knowledge of the likelihood function. Each requires only knowledge of
the likelihood peak and curvature (or variance) as well as a third
quantity related to the noise properties of the experiment. Alternately,
for already-calculated likelihood functions, each approximation gives a
functional form for fitting with a small number of parameters.

\subsection{The solution: approximating the likelihood}
\label{sec:solution}
\subsubsection{Offset lognormal distribution}
We already know enough about the likelihood to see a solution to this
problem. For a given multipole $\ell$, there are two distinct regimes of
likelihood. Consider a simple all-sky experiment with uniform pixel
noise and some symmetric beam. Taking such a map as our data, we write
$\bT_p = T_p + n_p$ which we transform to spherical harmonics, so the
data has contributions from the signal, $a_{\ell m}$, and the noise,
$n_{\ell m}$, and the likelihood is
\begin{equation}
  \label{eq:fullskylike}
-2\ln P(\bT|\C_{\ell})=
\sum_{\ell}
  (2\ell+1) \left[
    \ln\left(\C_\ell B_\ell^2 + \calN_\ell\right) 
    +{{\widehat \calD}_{\ell}\over\C_\ell B_\ell^2 + \calN_\ell}\right]
\end{equation}
( up to an irrelevant additive constant; cf.\ 
Eq.~\ref{eq:maplikelihood}), with
$\calN_\ell\equiv\ell(\ell+1)N_\ell/(2\pi)$, where
$N_\ell=\langle|n_{\ell m}|^2\rangle$ is the noise power spectrum in
spherical harmonics, and ${\widehat \calD}_{\ell} \equiv
[\ell(\ell+1)/(2\pi)]\ \sum_m |a_{\ell m}|^2/(2\ell+1)$ is the power
spectrum of the full data (noise plus beam-smoothed signal); we have
written it as a different symbol from above to emphasize the inclusion
of noise and again use script lettering to refer to quantities
multiplied by $\ell(\ell+1)/(2\pi)$.

Now, the likelihood is maximized at
$\C_\ell=({\widehat\calD}_\ell-\calN_\ell)/B_\ell^2$ and the curvature
about this maximum is given by
\begin{equation}\label{eq:fullcurv}
  {\cal F}^{(\C)}_{\ell\ell'}=
  -{\partial^2 \ln P(\bT|\C_{\ell})\over\partial \C_\ell\partial
    \C_{\ell'}} =
  {2\ell+1\over2}\left(\C_\ell+\calN_\ell/B_\ell^2 \right)^{-2}\delta_{\ell\ell'}
\end{equation}
so the error (defined by the variance) on a $\C_\ell$ is
\begin{equation}
  \delta \C_\ell = \left(\C_\ell+\calN_\ell/B_\ell^2 \right)/\sqrt{\ell+1/2}.
\end{equation}

Note that in this expression there is once again indication of a bias
if we assume Gaussianity: upward fluctuations have larger uncertainty
than downward fluctuations.  But this is not true for $Z_\ell$ where
$Z_\ell$ is defined so that $\delta Z_\ell \propto \delta \C_\ell /
(\C_\ell + \calN_\ell/B_\ell^2)$.  More precisely, $Z_\ell \equiv
\ln(\C_\ell + \calN_\ell/B_\ell^2)$.  Since $\delta Z_\ell$ is
proportional to a constant, our approximation to the likelihood is to
take $Z_\ell$ as normally distributed.  That is, we approximate
\begin{equation}\label{eq:gausslike}
 -2\ln P(\bT|\C_{\ell})=\sum_{\ell\ell'} Z_\ell
\left(M^{(Z)}\right)^{-1}_{\ell\ell'} Z_{\ell'}
\end{equation}
(up to a constant) where $M^{(Z)}_{\ell\ell'} = (\C_\ell+x_\ell)M^{(\C)}_{\ell\ell'}
(\C_{\ell'}+x_{\ell'})$ where $M^{(\C)}$ is the covariance matrix of the
$\C_\ell$, usually taken to be the inverse of the curvature matrix.
We refer to Eq.~\ref{eq:gausslike} as the offset lognormal distribution
of $\C_\ell$. 
Somewhat more generally we write
\begin{equation}\label{eq:Zl}
  Z_\ell = \ln(\C_\ell + x_\ell)
\end{equation}
for some constant $x_\ell$, which for the case in hand is
$x_\ell=\calN_\ell/B_\ell^2$.

It is illustrative to derive the quantity $Z_\ell$ in a somewhat more
abstract fashion. We wish to find a change of variables from $\C_\ell$
to $Z_\ell$ such that the curvature matrix is a constant:
\begin{equation}
  {\partial {\cal F}^{(Z)}_{\ell\ell'}\over\partial Z_\ell}=0.
\end{equation}
That is, we want to find a change of variables such that
\begin{equation}\label{eq:changevar}
  \left({\cal F}^{(Z)}\right)^{-1}_{LL'} = \sum_{\ell\ell'}
{\partial Z_L\over\partial \C_\ell} \left({\cal F}^{(\C)}\right)^{-1}_{\ell\ell'}
{\partial Z_{L'}\over\partial \C_{\ell'}}
\end{equation}
is not a function of $Z$. We immediately know of one such transformation
which would seem to do the trick:
\begin{equation}
\label{eq:dotrick}
  {\partial Z_L\over\partial \C_\ell} = {\cal F}_{L\ell}^{1/2}
\end{equation}
where the ${1/2}$ indicates a Cholesky decomposition or Hermitian square
root. In general, this will be a horrendously overdetermined set of
equations, $N^2$ equations in $N$ unknowns. However, we can solve this
equation in general if we take the curvature matrix to be
given everywhere by the diagonal
form for the simplified experiment we have been
discussing (Eq.~\ref{eq:fullcurv}).  In this case, the equations decouple and lose their
dependence on the data, becoming (up to a constant factor)
\begin{equation}
  {dZ_\ell\over d\C_\ell} = \left(\C_\ell+\calN_\ell/B_\ell^2\right)^{-1}.
\end{equation}
The solution to this differential equation is just what we expected,
\begin{equation}
  Z_\ell=\ln(\C_\ell+\calN_\ell/B_\ell^2)
\end{equation}
with correlation matrix
\begin{equation}
  \left({\cal F}^{(Z)}\right)^{-1}_{\ell\ell'}=
  {\left({\cal F}^{(\C)}\right)^{-1}_{LL'}\over
  \left(\C_L+\calN_L/B_L^2\right)\left(\C_{L'}+\calN_{L'}/B_{L'}^2\right)}
\end{equation}
where $\ell = L$ and $\ell'=L'$.
Please note that we are calculating a constant correlation matrix; the
$\C_\ell$ in the denominator of this expression should be taken at the
peak of the likelihood (i.e., the estimated quantities).

We emphasize that, even for an all-sky continuously and uniformly
sampled experiment (for which Eq.~\ref{eq:fullcurv} is exact), this
Gaussian form, Eq.~\ref{eq:gausslike}, is only an approximation, since
the curvature matrix is given by Eq.~\ref{eq:fullcurv} only at the peak.
Nonetheless we expect it to be a better approximation than a naive
Gaussian in $\C_\ell$ (which we note is the limit $x_\ell\to\infty$ of
the offset lognormal).



We have also considered another ansatz known as the equal variance
approximation which we have found sometimes reproduces the shape of the
likelihood better than the offset lognormal form above, especially in
the case of upper limits on $C_\ell$.

\subsection{Noise power}
In the above, we have derived the extra quantity needed in this ansatz,
$x_\ell=\calN_\ell$, but this was explicitly done only for the
unrealistic case of an experiment looking at the whole sky with uniform
noise, with a likelihood function Eq.~\ref{eq:fullskylike}. Can we
generalize this to more complicated experiments, which have the more
complicated likelihood function Eq.~\ref{eq:maplikelihood}, with
correlation matrices encoding such effects as
\begin{itemize}
\item nonuniform noise;
\item correlated noise;
\item incomplete sky coverage; and/or
\item complicated chopping schemes?
\end{itemize}
Happily, the answer to this question is that we can, and have
succesfully applied the results to experiments as diverse as COBE/DMR
and Saskatoon. 

For a map with noise correlation matrix $C_N$, and signal correlation
matrix $C_T$, we generalize our definition of $x$ to 
\begin{equation}
  \label{eq:getxb}
  \C_B/x_B = \sqrt{
    {\rm Tr}\left(C_N^{-1} C_{T,B} C_N^{-1} C_{T,B}\right)
    \over
    {\rm Tr}\left(C_T^{-1} C_{T,B} C_T^{-1} C_{T,B}\right)}
  \simeq
  \sqrt{
    {\rm Tr}\left(C_N^{-1} C_{T,B} C_N^{-1} C_{T,B}\right)
    \over
    {\rm Tr}\left(C^{-1} C_{T,B} C^{-1} C_{T,B}\right)} - 1
\end{equation}
where we use $B$ to label bands of $\ell$ rather than individual
multipoles, so that $\C_\ell$ has a fixed shape in each band with
amplitude $\C_B$ and $C_{T,B}=\partial C_T/\partial \C_B$ is the
derivative of the signal correlation matrix with respect to the power in
the band. The second expression would be appropriate for a full-sky
experiment where the signal and noise separate in the traces; it may be
numerically easier to calculate due to zero or negative eigenvalues in
$C_T$ that often appear due to approximation.  Other expressions that
may be more appropriate for particular experimental strategies are given
in Ref.~\cite{bjkii}.

\subsection{Tests and Results}

Because exact likelihoods have been calculated for many different
experiments to date, we can check this ansatz in great detail. (In the
future, this will probably be impossible because the exact calculation
of the likelihood is so expensive that other techniques for finding the
maximum will be necessary\cite{bcjk99}.)

\subsubsection{COBE/DMR}
\begin{figure}[htbp]
  \begin{center}\leavevmode
    \epsfxsize=5in\epsfbox{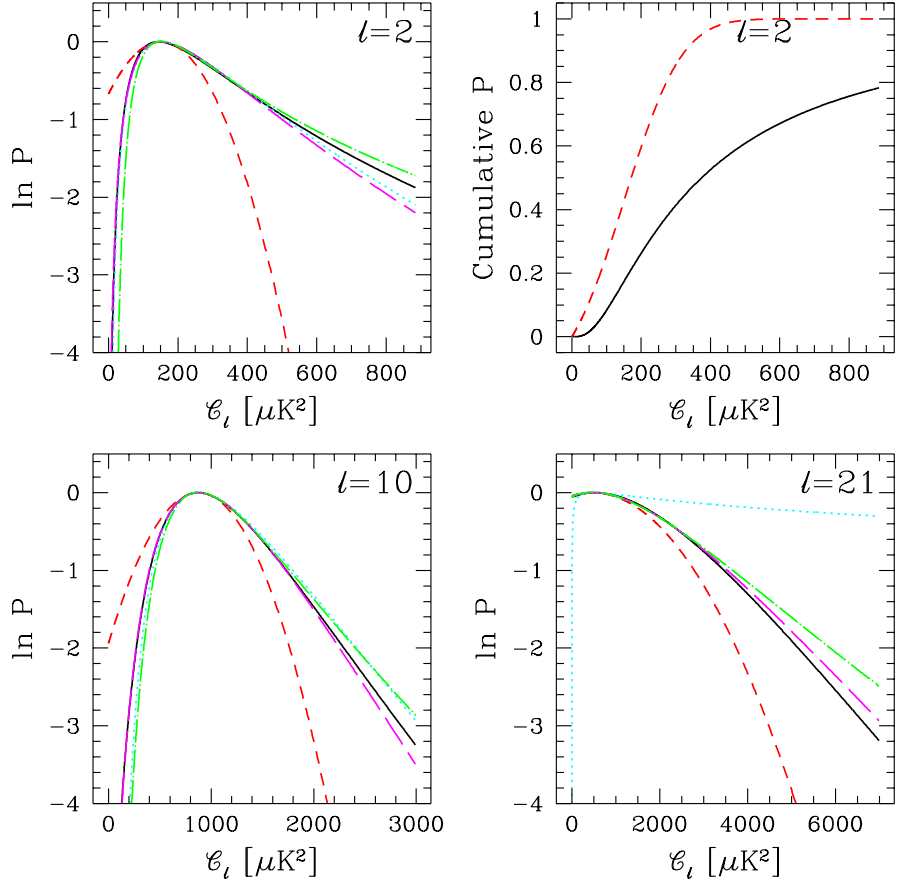}
    \caption[likefigs_2]{Full and approximate COBE/DMR likelihoods
    $P(\Delta|\C_\ell)$ for various values of $\ell$, as marked. The
    horizontal axis is ${\cal C}_\ell=\ell(\ell+1)C_\ell/(2\pi)$. The
    upper right panel gives the cumulative probability. The solid (black)
    line is the full likelihood calculated exactly. The short-dashed (red)
    line is the Gaussian approximation about the peak. The dotted (cyan)
    line is a Gaussian in $\ln{\C_\ell}$; the dashed (magenta) line is a
    Gaussian in $\ln{(\C_\ell+x_\ell)}$, as discussed in the text. The
    dot-dashed (green) line is the equal-variance approximation.
    \label{fig:dmrlike}}
  \end{center}
\end{figure}

First, we compare with the likelihood for COBE/DMR at several values of
$\ell$. In Fig.~\ref{fig:dmrlike} we show the actual likelihood for the
COBE quadrupole and other multipoles, along with the Gaussian that would
be assumed given the curvature matrix calculated from the data.  The
figures show another way of understanding the bias introduced by
assuming Gaussianity: upward deviations from the mean (which is not
actually the mean of the non-Gaussian distribution, but the mode) are
overly disfavored by the Gaussian distributions while downward ones are
overly probable.  For example, the standard-CDM value of
${\C}_2=770\muK^2$ is only 0.2 times less likely than the most likely
value of $150\muK^2$ but it seems like a 5-sigma excursion ($4\times
10^{-6}$ times less likely) based on the curvature alone.

Although it is extremely pronounced in the case of the quadrupole this
is a problem that plagues all CMB data: the actual distribution is
skewed to allow larger positive excursions than negative.  The full
likelihood ``knows'' about this and in fact takes it into account;
however, if we compress the data to observed $\C_\ell\pm\sigma_\ell$ (or
even observed $\C_\ell$ and a correlation matrix $M_{\ell\ell'}$) we
lose this information about the shape of the likelihood function.
Because of its relation to the well-known phenomenon of cosmic variance,
we choose to call this problem one of {\em cosmic bias}.

We emphasize that cosmic bias can be important even in high-S/N
experiments with many pixels. We might expect the central limit theorem
to hold in this case and the distributions to become Gaussian. Indeed
they do, at least near the peak. However, the central limit theorem does
not guarantee that the tails of the distribution will be Gaussian and
there is the danger that a few seemingly discrepant points are given
considerably more weight than they deserve. Cosmic bias has also been
noted in previous work \cite{bjki,uroscopy,OhSpergelHinshaw}.

Putting the problem a bit more formally, we see that even in the limit
of infinite signal-to-noise we cannot use a
simple $\chi^2$ test on $\C_\ell$ estimates; such a test implicitly
assumes a Gaussian likelihood. Unlike the distribution discussed here, a
Gaussian would have constant curvature ($\delta \C_\ell={\rm constant}$),
rather than $\delta \C_\ell\propto \C_\ell$ as illustrated here.

\subsubsection{Saskatoon}
We apply this ansatz to the Saskatoon experiment, perhaps the apotheosis
of a complicated chopping experiment.  The Saskatoon data are reported
as complicated chopping patterns ({\it i.e.}, beam patterns, $H$, above)
in a disk of radius about $8^\circ$ around the North Celestial Pole. The
data were taken over 1993-1995 (although we only use the 1994-1995 data)
at an angular resolution of $1.0$--$0.5^\circ$ FWHM at approximately
30~GHz and 40~GHz. More details can be found in
Refs.~\cite{nett95,woll95}.  The combination of the beam size, chopping
pattern, and sky coverage mean that Saskatoon is sensitive to the power
spectrum over the range $\ell=50$--$350$.  The Saskatoon dataset is
calibrated by observations of supernova remnant, Cassiopeia--A.\ Leitch
and collaborators\cite{ELthesis} have recently measured the flux and
find that the remnant is 5\% brighter than the previous best
determination.  We have renormalized the Saskatoon data accordingly.

\begin{figure*}[htbp]
  \begin{center}\leavevmode
    \epsfxsize=5in\epsfbox{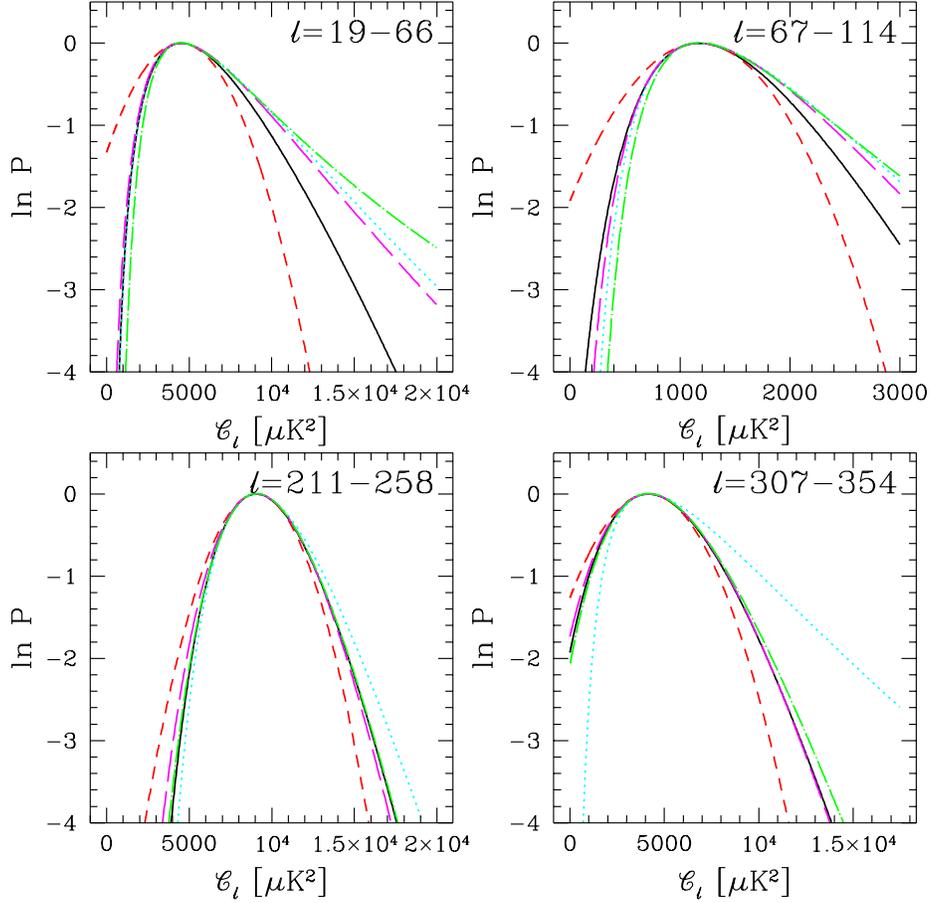}
  \caption{Full and approximate Saskatoon likelihoods. As in
    Fig.~\ref{fig:dmrlike}. The solid (black) line is the full
    likelihood calculated exactly. The short-dashed (red) line is the
    Gaussian approximation about the peak. The dotted (cyan) line is a
    Gaussian in $\ln{\C_\ell}$; the dashed (magenta) line is a Gaussian
    in $\ln{(\C_\ell+x_\ell)}$, as discussed in the text. The dot-dashed
    (green) line is the equal-variance approximation.
    \label{fig:sklike}}
\end{center}
\end{figure*}

We calculated $\C_\ell$ for this dataset in \cite{bjki}. We combine
these results with the data's noise matrix to calculate the appropriate
correlation matrixes (in this case, the full curvature matrix) for
Saskatoon and hence the appropriate $x_B$ (Eq.~\ref{eq:getxb}) and thus
our approximations to the full likelihood.  In Figure~\ref{fig:sklike},
we show the full likelihood, the naive Gaussian approximation, and our
present offset lognormal and equal-variance forms.  Again, both
approximations reproduce the features of the likelihood function
reasonably well, even into the tails of the distribution, certainly
better than the Gaussian approximation.  They seem to do considerably
better in the higher-$\ell$ bands; even in the lower $\ell$ bands,
however, the approximations result in a {\em wider}\/ distribution which
is preferable to the narrower Gaussian and its resultant strong bias.
Moreover, we have found that we are able to reproduce the shape of the
true likelihood essentially perfectly down to better than ``three
sigma'' if we simply {\em fit}\/ for the $x_B$ (but of course this can
only be done when we have already calculated the full
likelihood---precisely what we are trying to avoid!). For existing
likelihood calculations, this method can provide better results without
any new calculations (see \cite{bjkii} for our recommendations for the
reporting of CMB bandpower results for extant, ongoing, and future
experiments).

We have found that the shape of the power spectrum used with each bin
of $\ell$ can have an impact on the likelihood function
evaluated using this ansatz. Similarly, a finer binning in
$\ell$ will reproduce the full likelihood more accurately.
Although the maximum-likelihood amplitude
at a fixed shape ($n_s$) does not significantly depend on binning or
shape, the shape of the likelihood function along the
maximum-likelihood ridge changes with finer binning and with the assumed
spectral shape.


\subsubsection{COBE/DMR + Saskatoon}
\label{sec:results}
\begin{figure}[htbp]
  \begin{center}\leavevmode
    \epsfxsize=5in\epsfbox{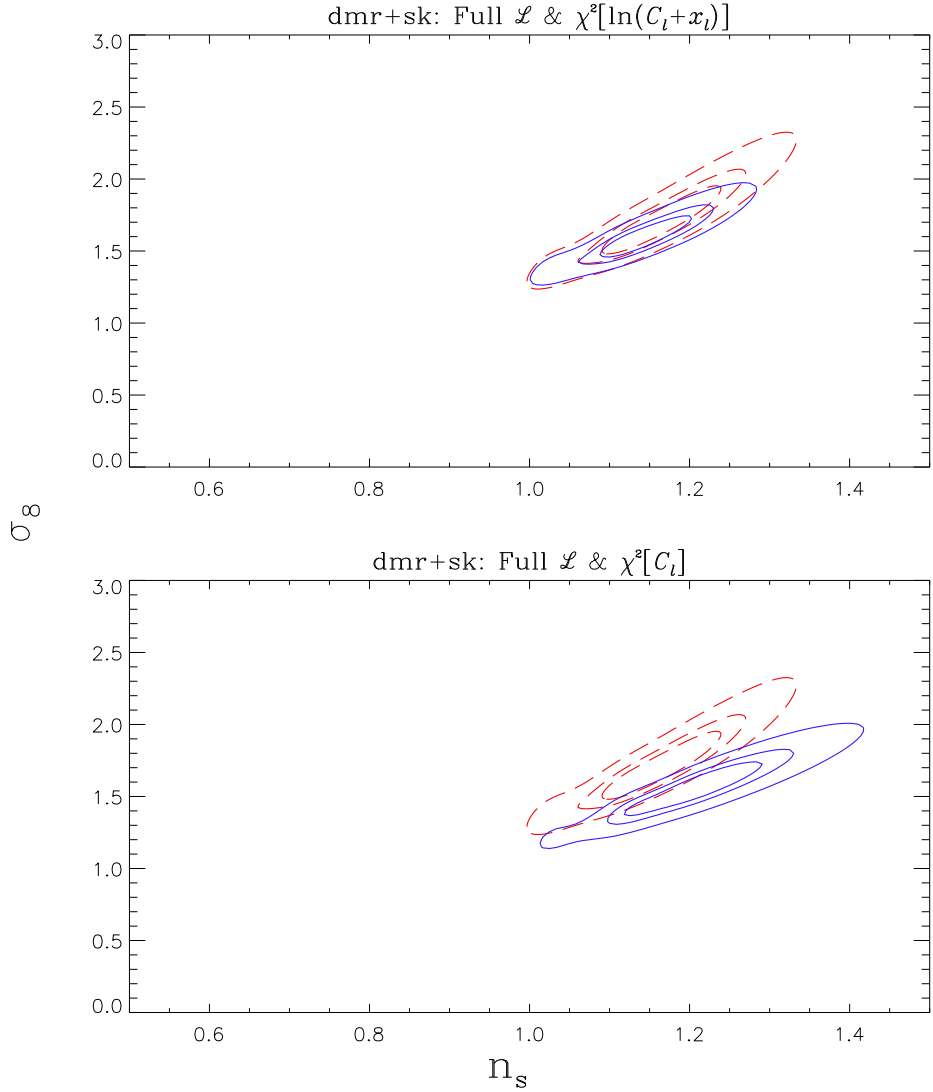}
  \caption{Likelihood contours for COBE/DMR and Saskatoon combined
    for the cosmological parameters $n_s$ and $\sigma_8$ (with otherwise
    standard CDM values) combining likelihoods
    from COBE/DMR and Saskatoon.  Contours are for ratios of the likelihood to
    its maximum equal to $\exp{-\nu^2/2}$ with $\nu=1,2,3$. Upper panel
    is for the full likelihood (dashed) and its offset lognormal
    approximation as a 
    Gaussian in $\ln{(\C_\ell+x_\ell)}$ (solid; see text); lower panel
    shows the full likelihood and its approximation as a Gaussian in
    $\C_\ell$.  
    \label{fig:dmrskcontours}}\end{center}
\end{figure}

As a further example and test of these methods, we can combine the
results from Saskatoon and COBE/DMR in order to determine cosmological
parameters. For this example, we use the orthogonal linear combinations
as described in the previous section.  In Figure~\ref{fig:dmrskcontours}
we show the likelihood contours for standard CDM, varying the scalar
slope $n_s$ and amplitude $\sigma_8$.  As before, we see that the naive
$\chi^2$ procedure is biased toward low amplitudes at fixed shape
($n_s$), but that our new approximation recovers the peak quite well.
The full likelihood gives a global maximum at $(n_s,
\sigma_8)=(1.15,1.67)$, and our approximation at $(1.13,1.58)$, while
the naive $\chi^2$ finds it at $(1.21,1.55)$, outside even the
three-sigma contours for the full likelihood. We can also marginalize
over either parameter, in which case the full likelihood gives
$n_s=1.17^{+0.08}_{-0.07}$, $\sigma_8=1.68^{+0.26}_{-0.21}$; our ansatz
gives $n_s=1.14^{+0.07}_{-0.05}$, $\sigma_8=1.60\pm0.15$; and the naive
$\chi^2$ gives $n_s=1.21^{+0.08}_{-0.09}$,
$\sigma_8=1.55^{+0.18}_{-0.20}$.
(Note that even with the naive $\chi^2$ we marginalize by explicit
integration, since the shape of the likelihood in parameter space is
non-Gaussian in all cases.)

\subsubsection{Parameter Estimation}
\label{sec:bands}
Above, we have discussed many different approximations to the likelihood
${\cal L}$.  Here we discuss finding the parameters that maximize this
likelihood (minimize the $\chi^2\equiv-2\ln{\cal L}$).  We then apply
our methods to estimating the power in discrete bins of $\ell$.  This
application provides another demonstration of the importance of using a
better approximation to the likelihood than a Gaussian.

The likelihood functions above depend on ${\C_\ell}$ which may
in turn depend on other parameters, $a_p$, which are, e.g.,
the physical parameters of a theory.  If we write the parameters as
$a_p +\delta a_p$ we can find the correction, $\delta a_p$,
that minimizes $\chi^2$ by solving
\begin{equation}
  \label{eq:quadest}
  \delta a_p = -\frac{1}{2}{\cal F}^{-1}_{pp'} { \partial \chi^2 \over
    \partial a_{p'}},
\end{equation}
where
\begin{equation}\label{eq:chi2curv}
{\cal F}_{pp'} \equiv \frac{1}{2}{ \partial^2 \chi^2 \over
\partial a_{p}\partial a_{p'} }
\end{equation}
is the curvature matrix for the parameters $a_p$.  If the $\chi^2$ were
quadratic (i.e., Gaussian ${\cal L}$) then Eq.~\ref{eq:quadest} would
be exact.  Otherwise, in most cases, near enough to its minimum,
$\chi^2$ is approximately quadratic and an iterative application of
Eq.~\ref{eq:quadest} converges quite rapidly.  The covariance matrix
for the uncertainty in the parameters is given by $\langle \delta a_p
\delta a_{p'} \rangle = {\cal F}^{-1}_{p p'}$. This is just an approximation to
the Newton-Raphson technique for finding the root of $\partial{\cal
  L}/\partial a_p = 0$; a similar techniqe is used in 
quadratic estimation of $\C_\ell$
\cite{tegmark,bjki,OhSpergelHinshaw}.

As our worked example here, we parameterize the power spectrum
by the power in $B=1$ to $11$ bins, $\C_B$.  Within each of the bins,
we assume $\C_\ell=\C_B$ to be independent of $\ell$.
We have chosen the offset lognormal approximation.  The $\chi^2$
completely describes the model:
\begin{eqnarray}
\label{eq:chisq}
\chi^2 & =& \sum_{i,j}\left(Z_i^{\rm t}-Z_i^{\rm d}\right)M_{ij}^Z
\left(Z_j^{\rm t}-Z_j^{\rm d}\right) + \chi^2_{\rm cal}; \\
\chi^2_{\rm cal} &\equiv & \sum_\alpha {(u_\alpha-1)^2
\over \sigma_\alpha^2}; \\
Z_i^{\rm d} & \equiv & \ln(D_i+x_i); \\
Z_i^{\rm t} &\equiv&
\ln\left(\sum_B u_{\alpha(i)}f_{iB}\C_B + x_i\right); \\
M_{ij}^Z & \equiv & M_{ij}\left(D_i+x_i\right)\left(D_j+x_j\right)\quad
\mbox{no sum};
\end{eqnarray}
where $M_{ij}$ is the weight matrix for the band powers
$D_i$.  We have
modeled the signal contribution to the data, $D_i$, as an average over
the power spectrum, $\sum_B f_{iB}\C_B$, times a calibration parameter,
$u_{\alpha(i)}$.  For simplicity, we take the prior probability
distribution for this parameter to be normally distributed.  Since the
datasets have already been calibrated, the mean of this distribution
is at $u_{\alpha}=1$.  The calibration parameter index, $\alpha$ is a
function of $i$ since different power spectrum constraints from the
same dataset all share the same calibration uncertainty.  We solve
simultaneously for the $u_{\alpha}$ and $\C_B$; i.e., together they
form the set of parameters, $a_p$, in Eq.~\ref{eq:quadest}.  For
those experiments reported as band-powers together with the trace of
the window function, $W^i_\ell$, the filter is taken to be
\begin{equation}
\label{eq:win2filt}
f_{iB}={\sum_{\ell \in B} W^i_\ell/\ell \over \sum_\ell W^i_\ell/\ell}.
\end{equation}
For Saskatoon and COBE/DMR, our $D_i$ are themselves estimates of the
power
in bands.  For these cases the above equation applies, but with
$W_\ell/\ell$ set to a constant within the estimated band and
zero outside.  The estimated bands have different $\ell$ ranges
than the target bands.

Instead of the curvature matrix of Eq.~\ref{eq:chi2curv} we use an
approximation to it that ignores a logarithmic term.  Including this
term can cause the curvature matrix to be non-positive definite as the
iteration proceeds.  The approximation has no influence on our
determination of the best fit power spectrum, but does affect the error
bars.  We expect that the effect is quite small.

We now proceed to find the best-fit power spectrum given different
assumptions about the value of $x_i$, binnings of the power spectrum and
editings of the data.  

\begin{figure}[htbp]
  \begin{center}\leavevmode
    \epsfxsize=5in\epsfbox{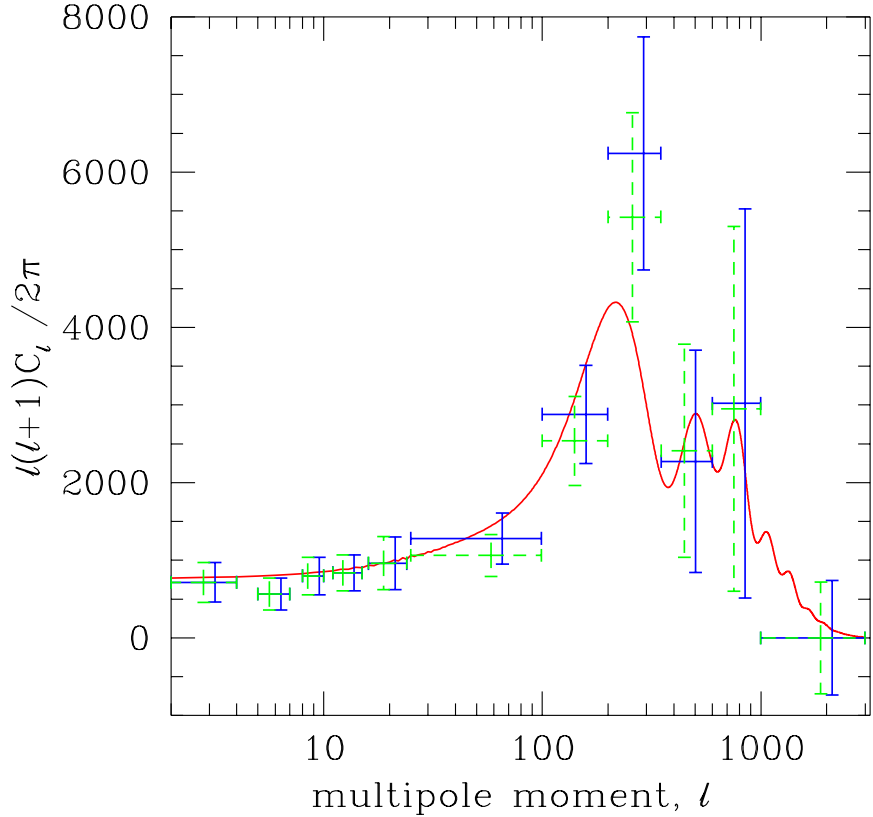}
  \caption{Power
    spectra that minimize the $\chi^2$ in Eq.~\ref{eq:chisq}.  The
    solid (dashed) error bars assume $x = 0$ ($x=\infty$) for those
    datasets with no determination of $x$; the two sets have been offset
    slightly for display purposes. Solid curve is standard CDM.
\label{fig:lnclvscl}}
\end{center}
\end{figure}

We have determined the $x_i$ only for COBE/DMR, Saskatoon, SP89, OVRO7
and SuZIE.  To test the
sensitivity to the unknown $x_i$s we found the minimum-$\chi^2$
power spectrum assuming the two extremes of $x_i=0$ (corresponding to
lognormal) and $x_i=\infty$ (corresponding to Gaussian).  These two
power spectra are shown in Fig.~\ref{fig:lnclvscl}.  Note that both
power spectra were derived using our measured $x_i$ values; only the
unknown $x_i$ values were varied.  
The variation in the results would be much greater if we let
these $x_i$ values be at their extremes. 

\section{Conclusions}

In this proceedings, we have discussed the beginning and end of the
CMB data-analysis process, and shown how they can be unified in a
Bayesian likelihood formalism. We have introduced new techniques and
approximations that
\begin{itemize}
\item allow simultaneous calculation of the noise properties and
  underlying sky map.
\item approximate the full cosmological likelihood function with minimal 
  information.
\end{itemize}

Crucially, both of these innovations can be calculated in relatively
small or comparable amounts of time to the most
computationally-challenging aspects of the problem. Hence, they await
the same innovations necessary to perform these calculations in the
first place for upcoming megapixel datasets\cite{bcjk99}.

\section*{Acknowledgements} 
AHJ would like to thank Mike Seifert, Julian Borrill, the COMBAT
collaboration, and the Saskatoon, MAXIMA and BOOMERANG teams for many
helpful conversations during the course of this research. AHJ
acknowledges support by NASA grants NAG5-6552 and NAG5-3941 and by NSF
cooperative agreement AST-9120005.

\end{document}